\documentstyle[12pt]{article}
\makeatletter

\@addtoreset{equation}{section}
\makeatother
\newcommand{\be}{\begin{equation}}
\newcommand{\ee}{\end{equation}}
\newcommand{\bea}{\begin{eqnarray}}
\newcommand{\eea}{\end{eqnarray}}
\newcommand{\KM}{Ka\v{c}-Moody}
\newcommand{\ZN}{${\bf Z}_N$}

\begin{document}

\begin{titlepage}
\null

\begin{flushright} KOBE-TH-95-04\\ August 1995 \end{flushright}

\vspace{1cm}
\begin{center}

{\Large\bf
Asymmetric Orbifold Models of\\
Non-supersymmetric Heterotic Strings\\
}

\vspace{2.5cm}
{\large
Toshihiro Sasada\\
}

\vspace{5mm}
\baselineskip=7mm
{\sl
Graduate School of Science and Technology\\
Kobe University\\
Rokkodai, Nada, Kobe 657, Japan\\
}

\vspace{3cm}
{\large\bf Abstract\\
}

\end{center}

We investigate asymmetric orbifold models constructed from non-supersymmetric
heterotic strings.
We systematically classify the asymmetric orbifold models with standard
embeddings and present a list of asymmetric orbifolds which are geometrically
interpreted as toroidal compactifications of non-supersymmetric heterotic
strings.
By studying non-standard embedding models, we also construct examples of the
{\em supersymmetric} asymmetric orbifold models based on non-supersymmetric
heterotic strings.

\end{titlepage}

\baselineskip=7mm

\section{Introduction}

Asymmetric orbifolds \cite{NSV} belong to the efficient construction schemes
of 4D string models.
In constructing asymmetric orbifolds, we consider the left-right asymmetric
twists of the momentum lattices of toroidal compactifications.
Thus, we have in general no direct geometric interpretation of asymmetric
orbifolds.
However, some class of asymmetric orbifolds may possess the geometric
interpretation through the quantum equivalence to other compactifications.
In the previous paper \cite{Fermion}, we have studied the geometric
interpretation of asymmetric orbifolds as Narain's toroidal
compactifications~\cite{NSW} and found a simple condition for the geometric
interpretation of asymmetric orbifolds.

In this paper we construct the asymmetric orbifold models which are
based on the non-supersymmetric heterotic strings \cite{HETERO},
and investigate the geometric interpretation of such asymmetric orbifolds as
the toroidal compactification \cite{TORUS} of non-supersymmetric heterotic
strings.
Asymmetric orbifold models of non-supersymmetric heterotic strings have not
been studied before.
The main purpose of this paper is to extend the previous analysis by studying
the geometric interpretation of the {\em non-supersymmetric} asymmetric
orbifold models.
We classify the asymmetric orbifold models with standard
embeddings in a systematic way and give a list of the asymmetric orbifold
models which are geometrically interpreted as toroidal compactifications
of non-supersymmetric heterotic strings.

We shall study the condition for interpreting orbifold models as toroidally
compactified ones.
A crucial feature of ten-dimensional heterotic strings is the existence
of NSR fermions.
They will form an $SO(8)$ \KM\ algebra in the light-cone gauge.
Since the compactification onto tori yields no effect on the NSR fermions,
the fermions on orbifolds should generate $SO(8)$ \KM\ algabras if they are
interpreted as toroidal compactifications.
We define fermion currents as the current operators with conformal weight one
which generate $SO(N)$ algebra and which contain the $SO(2)$ \KM\ algebra
generated by the space-time component of the fermions on orbifold models.
A necessary condition for interpreting orbifold models as toroidally
compactified ones is that the fermion currrents should generate $SO(8)$ \KM\
algebras.
We shall show that the above condition is necessary and also sufficient
for our class of asymmetric orbifolds.
We should note that symmetric orbifolds cannot be interpreted as toroidal
compactifications since fermion currents on symmetric orbifolds are
constructed by the currents in the untwisted sector which are always smaller
than $SO(8)$.
Orbifold models must possess twist-untwist intertwining currents
\cite{CHDGM,ISST} in order to possess $SO(8)$ fermion currents.
Thereby, orbifold models must be asymmetric if they are interpreted as
toroidal compactifications.

The above asymmetric orbifold models with standard embeddings possess no
supersymmetry.
At first sight, any orbifold model constructed from non-supersymmetric
heterotic
strings seems to possess no supersymmetry.
However, this will turn out not to be true for generic asymmetric orbifold
models.
In fact, for asymmetric orbifold models of supersymmetric heterotic strings
supersymmetry can appear from the twisted sectors~\cite{ISST,I}.
We may expect that the similar mechanism still holds for some classes of the
asymmetric orbifolds constructed from non-supersymmetric heterotic strings.
In this paper, by studying non-standard embedding models, we shall show some
examples of {\em supersymmetric} asymmetric orbifold models
constructed from non-supersymmetric heterotic strings.

The organization of this paper is as follows.
In section 2, we explain the construction of the asymmetric orbifolds from
non-supersymmetric heterotic strings.
In section 3, we clarify a necessary and sufficient condition for the
asymmetric orbifold models to be interpreted as toroidally compactified
models.
In section 4, we classify the asymmetric orbifold models and investigate
fermion currents on them.
In section 5, we construct the examples of supersymmetric asymmetric
orbifolds of non-supersymmetric heterotic strings.
In section 6, we present our conclusion.

\section{Asymmetric orbifold models}

We shall recall the basic set up of the non-supersymmetric
ten-dimensional heterotic strings \cite{HETERO}.
The non-supersymmetric heterotic strings with rank 16 gauge groups
are specified by the left-moving 16-dimensional conjugacy classes
$\Gamma_{i}^{16,0}$ $(i = 1, \dots, 4)$.
The conjugacy class $\Gamma_{1}^{16,0}$ $(\Gamma_{2}^{16,0})$ is
connected to the NS sector states with the even (odd) G-parity.
The conjugacy class $\Gamma_{3}^{16,0}$ $(\Gamma_{4}^{16,0})$ is
connected to the R sector states with the positive (negative) chirality.
For definiteness, we will concentrate on the tachyon-free non-supersymmetric
heterotic strings with $SO(16) \times SO(16)$ gauge groups.
Then $\Gamma_{i}^{16,0}$ $(i = 1, \dots, 4)$ are given by
$\Gamma_{1}^{16,0} = (0, 0) \cup (c, c)$,
$\Gamma_{2}^{16,0} = (s, v) \cup (v, s)$,
$\Gamma_{3}^{16,0} = (v, v) \cup (s, s)$ and
$\Gamma_{4}^{16,0} = (c, 0) \cup (0, c)$,
where $0$, $v$, $s$ and $c$ are the adjoint, vector, spinor and conjugate
spinor conjugacy classes of $SO(16)$, respectively.
Almost arguments in this paper will be straightforwardly applied to the other
non-supersymmetric heterotic strings.

Let us now construct asymmetric \ZN -orbifolds of the non-supersymmetric
heterotic string.
We shall start with the toroidal compactification \cite{TORUS} of the
non-supersymmetric heterotic string defined by the conjugacy classes
$\Gamma_{i}^{16,0} \oplus \Gamma^{6,6}$ $(i = 1, \dots, 4)$, where
$\Gamma^{6,6}$ is a (6+6)-dimensional Lorentzian even self-dual lattice.
The left- and right-moving momentum $(p_L^I, p_L^i, p_R^i)$
$(I = 1, \dots, 16; i = 1, \dots, 6)$ lies on the conjugacy classes
$\Gamma_{i}^{16,0} \oplus \Gamma^{6,6}$.
The group element~$g$ which generates a cyclic group \ZN\ is defined by the
following action on the string coordinates:
\be
g: (X_L^I, X_L^i, X_R^i) \rightarrow (X_L^I + 2\pi v_L^I, U_L^{ij} X_L^j,
U_R^{ij} X_R^j),
\ee
where $U_L$ and $U_R$ are rotation matrices which satisfy
$U_L^N = U_R^N = {\bf 1}$ and $v_L^I$ is a shift vector.
The rotation matrices $U_L$ and $U_R$ must be an automorphism of
$\Gamma^{6,6}$:
\be
(U_L^{ij} p_L^j, U_R^{ij} p_R^j) \in \Gamma^{6,6} \ \  {\rm for \ all} \
(p_L^i,p_R^i) \in \Gamma^{6,6}.
\ee
The action of the operator $g$ on the right-moving fermions is given by the
$U_R$ rotation.
Let $N_{\ell}$ be the minimum positive integer such that
$(g^{\ell})^{N_{\ell}} = 1$ in the $g^{\ell}$-twisted sector.
We shall denote the eigenvalues of $U_L^{\ell}$ and $U_R$ by
$ \{ e^{i 2\pi \zeta_{\ell}^a}, e^{-i 2\pi \zeta_{\ell}^a} ; a=1,2,3 \} $
and
$ \{ e^{i 2\pi \zeta_R^a}, e^{-i 2\pi \zeta_R^a}; a=1,2,3 \} $,
respectively.
Then we have the level matching condition for the one-loop modular invariance
for $N_{\ell}$ odd
\be
N_{\ell}
\left[
  \frac{1}{2} (\ell v_L^I)^2
+ \frac{1}{2} \sum_{a=1}^{3} \zeta_{\ell}^a (1-\zeta_{\ell}^a)
\right]
= 0 \ \ \bmod 1,
\ee
\be
N_{\ell} \sum_{a=1}^3 \ell \zeta_R^a = 0 \ \  \bmod 2;
\ee
for $N_{\ell}$ even, in addition to the above conditions, we have
\be
  p_L^i ( U_L^{\ell\frac{N_{\ell}}{2}} )^{ij} p_L^j
- p_R^i ( U_R^{\ell\frac{N_{\ell}}{2}} )^{ij} p_R^j = 0
\ \ \bmod 2
\ee
for all $(p_L^i,p_R^i) \in \Gamma^{6,6}$.

\section{Torus-orbifold equivalence of heterotic strings}

We now show that the necessary condition for the equivalence with torus
compactifications in terms of the fermion currents is also a sufficient
condition for our class of asymmetric orbifold models.
As we shall see in the next section, the \ZN -transformation of our
asymmetric orbifold model is an inner automorphism of the momentum lattice.
Then, the \ZN -transformation on the lattice $\Gamma^{6,6}$ is equivalent to
a shift \cite{NSV,ISST,DHVW,S}.
We shall use the bosonized representation of world-sheet
fermions.
The momentum $p_R^t$ $(t = 1, \dots, 4)$ of the fermions lies on the weight
lattice of $SO(8)$.
The momentum in the vector (adjoint) conjugacy class corresponds to the state
in the NS sector with even (odd) G-parity.
The momentum in the spinor (conjugate spinor) conjugacy class corresponds to
the state in the R sector with positive (negative) chirality.
The \ZN -transformation $g$ acts on the bosons as a shift, where the
shift vector is given by $v_R^t = (\zeta_R^a, 0)$.
Let us denote the conjugacy classes of $SO(2n)$ as $(i)_n$
$(i = 0, v, s, c)$.
The momentum
$(p_L^{\prime I}, p_L^{\prime i}, p_R^{\prime i}, p_R^{\prime t})$
$(I = 1, \dots, 16; i = 1, \dots, 6; t = 1, \dots, 4)$
in the $g^{\ell}$-sector
($\ell = 0$ for untwisted sector and $\ell = 1, \dots, N-1$ for twisted
sectors)
of the asymmetric orbifolds lies on the following lattice:
\be
[
( \Gamma_{1}^{16,0} \oplus \Gamma^{6,6} \oplus (v)_4 )
\cup
( \Gamma_{2}^{16,0} \oplus \Gamma^{6,6} \oplus (0)_4 )
]
+ \ell (v_L^I, v_L^i, v_R^i, v_R^t),
\ee
for NS sector, and
\be
[
( \Gamma_{3}^{16,0} \oplus \Gamma^{6,6} \oplus (s)_4 )
\cup
( \Gamma_{4}^{16,0} \oplus \Gamma^{6,6} \oplus (c)_4 )
]
+ \ell (v_L^I, v_L^i, v_R^i, v_R^t),
\ee
for R sector.
The operator $g$ in the $g^{\ell}$-sector
will be expressed as
\be
g = \eta_{\ell} \exp [ i 2\pi (p_L^{\prime I} v_L^I + p_L^{\prime i}
v_L^i - p_R^{\prime i} v_R^i - p_R^{\prime t} v_R^t ) ],
\ee
where $\eta_{\ell}$ is a constant phase and $(v_L^i, v_R^i)$ is a
shift vector which satisfy $N (v_L^i, v_R^i) \in \Gamma^{6,6}$.
The phase $\eta_{\ell}$ is determined from the modular transformations
\cite{ISST}:
\be
\eta_{\ell}
= \exp \{ -i \pi \ell [ (v_L^I)^2 + (v_L^i)^2 - (v_R^i)^2 - (v_R^t)^2 ] \}.
\ee
Every physical state  in the $g^{\ell}$-sector must satisfy
the condition $g = 1$ because it must be invariant under the
\ZN -transformation.

In order to examine \KM\ algebras on heterotic string models, it
may be convenient to use the bosonic string map \cite{LSW} and
investigate \KM\ algebras on the corresponding bosonic string
models.
We first decompose the momentum lattices of the asymmetric orbifolds with
respect to $SO(2)$ conjugacy classes to which the momentum $p_R^{\prime t}$
$( t=4 )$ belongs.
Then, preserving the modular transformation properties, the $SO(2)$ conjugacy
classes $(i)_1$ $(i = 0, v, s, c)$ are mapped to the $SO(10) \times E_8$
conjugacy classes $(i)_5 \oplus \Gamma^{0,8}$ $(i = 0, v, s, c)$ as follows:
$(0)_1 \rightarrow (v)_5 \oplus \Gamma^{0,8}$,
$(v)_1 \rightarrow (0)_5 \oplus \Gamma^{0,8}$,
$(s)_1 \rightarrow (s)_5 \oplus \Gamma^{0,8}$ and
$(c)_1 \rightarrow (c)_5 \oplus \Gamma^{0,8}$,
where $\Gamma^{0,8}$ is a root lattice of $E_8$.
After the bosonic string map, the momentum
$(p_L^{\prime I}, p_L^{\prime i}, p_R^{\prime i}, p_R^{\prime t})$
in the $g^{\ell}$-sector lies on the following lattice:
\bea
&&
[
( \Gamma_{1}^{16,0} \oplus \Gamma^{6,6} \oplus (0)_8 \oplus \Gamma^{0,8} )
\cup
( \Gamma_{2}^{16,0} \oplus \Gamma^{6,6} \oplus (v)_8 \oplus \Gamma^{0,8} )
\\
&\cup&
( \Gamma_{3}^{16,0} \oplus \Gamma^{6,6} \oplus (s)_8 \oplus \Gamma^{0,8} )
\cup
( \Gamma_{4}^{16,0} \oplus \Gamma^{6,6} \oplus (c)_8 \oplus \Gamma^{0,8} )
]
+ \ell (v_L^I, v_L^i, v_R^i, v_R^t),
\nonumber
\eea
where $p_R^{\prime t}$ $( t=4, \dots, 8; 9, \dots, 16)$ are defined as the
momentum which belong to the $SO(10) \times E_8$ conjugacy classes and
$v_R^t = (\zeta_R^a, 0^5; 0^8)$.
Therefore, on the corresponding bosonic strings, the momentum
$(p_L^{\prime I}, p_L^{\prime i}, p_R^{\prime i}, p_R^{\prime t})$
of the physical states in the $g^{\ell}$-sector lies
on a $(22+22)$-dimensional Lorentzian even self-dual lattice
$\Gamma^{\prime 22,22}$, where
$(p_L^{\prime I}, p_L^{\prime i}, p_R^{\prime i}, p_R^{\prime t})$
lies on the above lattice and satisfies the physical state condition:
\be
p_L^{\prime I} v_L^I + p_L^{\prime i} v_L^{\prime i}
- p_R^{\prime i} v_R^{\prime i} - p_R^{\prime t} v_R^t
- \frac{1}{2} \ell [ (v_L^I)^2 + (v_L^i)^2
- (v_R^i)^2 - (v_R^t)^2 ] = 0 \ \ \bmod 1.
\ee
If the fermion currents on asymmetric orbifolds generate
$SO(8)$ \KM\ algebras, then the corresponding $(22+22)$-dimensional
lattice $\Gamma^{\prime 22,22}$ is decomposed as follows:
\be
\Gamma^{\prime 22,22} =
[
( \Gamma_{1}^{\prime 22,6} \oplus (0)_8 )
\cup
( \Gamma_{2}^{\prime 22,6} \oplus (v)_8 )
\cup
( \Gamma_{3}^{\prime 22,6} \oplus (s)_8 )
\cup
( \Gamma_{4}^{\prime 22,6} \oplus (c)_8 )
]
\oplus \Gamma^{0,8},
\ee
where $\Gamma_{i}^{\prime 22,6}$ $(i = 1, \dots, 4)$ are
$(22+6)$-dimensional conjugacy classes.
This implies that, after reversing the bosonic string map,
we obtain the toroidal compactifications of the ten-dimensional
non-supersymmetric heterotic strings.

\section{Classification of asymmetric orbifold models}

Let us discuss the classification of asymmetric orbifold models.
Since we are investigating the geometric interpretation of asymmetric
orbifold models as toroidal compactifications, we classify the asymmetric
orbifolds with the right-moving twist-untwist intertwining currents.
We first consider the choice of the momentum lattices.
Unlike symmetric orbifolds models, the momentum lattices $\Gamma^{6,6}$ are
severely restricted by the left-right asymmetric automorphisms.
One of the known classes of such momentum lattices are given by
\be
\Gamma^{6,6} = \{(p_L^i, p_R^i) \vert
p_L^i, p_R^i \in \Lambda_W \ {\rm and}\  p_L^i - p_R^i \in \Lambda_R\},
\ee
where $\Lambda_W$ and $\Lambda_R$ are the weight and root lattices of a
simply-laced semisimple Lie algebra with the squared length of roots
normalized to two \cite{EN}.
The left- and right-moving rotation matrices $U_L$ and $U_R$ are taken to be
the Weyl group elements \cite{C} of the Lie algebra.
Then the matrices $U_L$ and $U_R$ always satisfy the condition
for the automorphism of $\Gamma^{6,6}$.
Next, we consider the standard embeddings in the gauge degrees of freedom.
Since the gauge group of the tachyon free non-supersymmetric strings is
$SO(16) \times SO(16)$, we can embed the shift $v_L^I$ in the $SO(6)$
subgroup of the first $SO(16)$.
Let us denote the eigenvalues of $U_L$ by $ \{ e^{i 2\pi \zeta_L^a},e^{-i
2\pi \zeta_L^a} ; a=1,2,3 \} $.
Then the shift vector $v_L^I$ is chosen as
\be
v_L^I = (\zeta_L^a, 0^5; 0^8).
\ee
By the above choice of the shift vector, the first equation of the level
matching conditions reduces to
\be
N_{\ell} \sum_{a=1}^3 \ell \zeta_L^a = 0 \ \  \bmod 2.
\ee
With these level matching conditions, we classify the modular invariant
asymmetric orbifold models.

To determine fermion currents on asymmetric orbifolds, we use
the bosonic string map.
We must investigate the full right-moving \KM\ algebras of
asymmetric orbifolds.
This is because, unlike symmetric orbifold case \cite{HMKKKOOT},
there is no simple diagrammatical method for determining \KM\
algebras if there exist twist-untwist intertwining currents.
The results of calculation are summarized in table 1.
We present in table 1 the right-moving \KM\ algebras
and the fermion currents of the asymmetric orbifold models.
As we have seen above, the necessary and sufficient condition for the
asymmetric orbifold models to be interpreted as toroidal compactifications of
non-supersymmetric heterotic strings is that the fermion currents on the
asymmetric orbifolds should generate $SO(8)$ \KM\ algebras.
In table 1, we see that many asymmetric orbifold models are geometrically
interpreted as toroidal compactifications of non-supersymmetric heterotic
strings.

\section{Supersymmetric asymmetric orbifold models}

We will present examples of the supersymmetric asymmetric orbifold models
constructed from non-supersymmetric heterotic strings.
We first consider the condition for obtaining massless gravitino states in
the
$g^{\ell}$-sector $(\ell = 0, 1, \dots, N-1)$ of asymmetric orbifold models.
The existence of a massless gravitino state will lead to the existence of a
supersymmetry.
Let us define the $g^{\ell}$-invariant sublattice $I_{\ell}$ of
$\Gamma^{6,6}$ by
\be
I_{\ell} = \{ (p^i_L,p^i_R) \in {\Gamma^{6,6}}\vert
( (U_L^{\ell})^{ij}p_L^j, (U_R^{\ell})^{ij}p_R^j ) = ( p_L^i, p_R^i ) \}.
\ee
The momentum $(p_L^i, p_R^i)$ of the $g^{\ell}$-twisted sector lies on the
lattice $I_{\ell}^{\ast} $, where $I_{\ell}^{\ast} $ is the dual
lattice of $I_{\ell}$.
The number of degeneracy of the ground states in the $g^{\ell}$-twisted
sector is given by
\be
n_{\ell} =
\frac{
\sqrt{ \det^{\prime} ({\bf 1} - U_L^{\ell})
\det^{\prime} ({\bf 1} - U_R^{\ell}) }
}{
{\rm vol}(I_{\ell})
},
\ee
where $\det^{\prime}$ is evaluated over the eigenvalues of $U_L^{\ell}$ and
$U_R^{\ell}$ not equal to one and ${\rm vol}(I_{\ell})$ is the volume of
the unit cell of the lattice $I_{\ell}$.
We shall denote the eigenvalues of $U_R^{\ell}$ by
$ \{ e^{i 2\pi {\bar \zeta}_{\ell}^a}, e^{-i 2\pi {\bar \zeta}_{\ell}^a};
 a=1,2,3 \} $.
The mass formula in the $g^{\ell}$-sector is given by
\bea
\frac{1}{8} m_L^2
& = &
\frac{1}{2} \sum_{I=1}^{16} (p_L^I + \ell v_L^I)^2
+ \frac{1}{2} \sum_{i=1}^{6}  (p_L^i)^2
+ \frac{1}{2} \sum_{a=1}^{3} \zeta_{\ell}^a (1-\zeta_{\ell}^a)
+ N_L
- 1,
\\
\frac{1}{8} m_R^2
& = & \frac{1}{2} \sum_{i=1}^{6} (p_R^i)^2
+ \frac{1}{2} \sum_{t=1}^{4} (p_R^t + \ell v_R^t)^2
+ \frac{1}{2} \sum_{a=1}^{3} {\bar \zeta}_{\ell}^a (1-{\bar
\zeta}_{\ell}^a)
+ N_R
- \frac{1}{2},
\eea
where $(p_L^i, p_R^i) \in I_{\ell}^{\ast}$,
$(p_L^I, p_R^t) \in
( \Gamma_{1}^{16,0}\oplus (v)_4 )
\cup
( \Gamma_{2}^{16,0}\oplus (0)_4 )
\cup
( \Gamma_{3}^{16,0}\oplus (s)_4 )
\cup
( \Gamma_{4}^{16,0}\oplus (c)_4 )$, and
$N_L$ and $N_R$ are the number operators of oscillators.
In order to have massless spin 3/2 states, we must have
$p_R^{t = 4} = \pm 1/2$, $N_L=1$, $U_L^{\ell} = {\bf 1}$ and
$p_L^{\prime I} \equiv p_L^I + \ell v_L^I = 0$ for
$p_L^I \in \Gamma^{16,0}_{3} \cup \Gamma^{16,0}_{4}$,
where the condition  $N_L=1$ must be satisfied by the oscillators of the
space-time coordinates.
It should be noted that
$p_L^{\prime I} = 0$ for
$p_L^I \in \Gamma^{16,0}_{3} \cup \Gamma^{16,0}_{4}$
never holds if we set
$\ell = 0$ (untwisted sector) or $v_L^I = 0$ (no embedding).
Let us denote the order of the left-moving shift (twist) as $N_S$
($N_L$).
Then, from the mass formula we can check that there is no massless gravitino
state for
the orbifold models with $N_S = N_L$.
Examples of the models satisfying such a condition are the orbifold models
with standard embeddings and symmetric orbifolds.
Thus, we have to consider the asymmetric orbifold models with
non-standard embeddings in order to obtain massless gravitino states.

We now construct examples of supersymmetric asymmetric orbifold models.
We start with a toroidal compactification of $SO(16)\times SO(16)$
non-supersymmetric heterotic strings, where the momentum lattice
$\Gamma^{6,6}$ is associated with a Lie algebra $(SU(3))^3$.
The left- and right-twists of asymmetric orbifolds are taken to be the
${\bf Z}_3$-twist matrix $U$ whose eigenvalues are given by
$\{e^{2 \pi i \zeta^a}, e^{-2 \pi i \zeta^a}; a = 1, 2, 3\}$ with
$\zeta^a = (1/3, 1/3, 2/3)$.
We take the shift vector to be $v_L^I \in \Gamma_{3}^{16,0}$.
For example, such shift vectors are given by
$v_L^I = (1, 0^7; 1, 0^7)$ and $v_L^I = ((1/2)^8; (1/2)^8)$.
The order of the shift is given by $N_S = 2$, and the level matching
conditions are satisfied since $(v_L^I)^2 = 0$ $\bmod$ $2$.
The first example is the asymmetric ${\bf Z}_6$-orbifold model with
$U_L = {\bf 1}$ and $U_R = U$.
Massless gravitino states may appear from $\ell = 1, 3, 5$ twisted sectors
since
the solution of $p_L^{\prime I} = 0$ exists for
$\ell = 1, 3, 5$ if we set $p_L^I \in \Gamma_{3}^{16,0}$.
In the $\ell = 1$ sector, from the massless condition we obtain
$p_R^t = (-1/2, -1/2, -1/2, -1/2)$.
The degeneracy of the ground states is given by $n_1 = 1$.
This state is physical since all massless states in the first twisted sector
are known to be physical \cite{FIQS}.
In the $\ell = 3$ sector, we obtain
$p_R^{\prime t} = (\pm 1/2, \pm 1/2, \pm 1/2, \pm 1/2)$,
where the number of minus signs should be even.
Since $U_L^3 = U_R^3 = {\bf 1}$, the $g$ operator in this sector will be
given by $g = \exp [ p_L^{\prime I} v_L^I - p_R^{\prime t} v_R^t
-\frac{1}{2} \ell ( (v_L^I)^2 - (v_R^t)^2) ]$.
Thus, physical (i.e. $g = 1$) states are given by
$p_R^{\prime t} = \pm (1/2, 1/2, -1/2, -1/2)$.
In the $\ell = 5$ sector, we obtain $p_R^t = (-3/2, -3/2, -7/2, 1/2)$.
The degeneracy of the ground states is given by $n_5 = 1$.
This state is physical since all massless states in the
$g^5$-twisted sector are physical.
The existence of these massless states will lead to the existence of two
gravitino states.
Therefore, in this model we have $N=2$ supersymmetry.
The second example is the asymmetric ${\bf Z}_6$-orbifold model with
$U_L = U$ and $U_R = {\bf 1}$.
{}From the mass formula, we see that gravitino states may appear from
the $\ell =3$ twisted sector.
In the $\ell = 3$ sector, we obtain
$p_R^{\prime t} = (\pm 1/2, \pm 1/2, \pm 1/2, \pm 1/2)$, where the number of
minus signs should be even.
The $g$ operator takes the value $g = 1$ for all such massless states.
Thus, the above states are all physical.
The existence of these massless states will lead to the existence of four
gravitino states.
Therefore, in this model we have $N=4$ supersymmetry.
This model will possess the geometric interpretation as a Narain's toroidal
compactification by the mechanism discussed in the previous
paper~\cite{Fermion}.
The last example is the asymmetric ${\bf Z}_6$-orbifold model with
$U_L  = U_R = U$.
We see that gravitino states may appear from the $\ell =3$ twisted sector.
In the $\ell = 3$ sector, we obtain
$p_R^{\prime t} = (\pm 1/2, \pm 1/2, \pm 1/2, \pm 1/2)$, where the number of
minus signs should be even.
The $g$ operator will be given by
$g = \exp [ p_L^{\prime I} v_L^I - p_R^{\prime t} v_R^t
-\frac{1}{2} \ell ( (v_L^I)^2 - (v_R^t)^2) ]$.
Then physical states are given by
$p_R^{\prime t} = \pm (1/2, 1/2, -1/2, -1/2)$.
The existence of these massless states will lead to the existence of one
gravitino state.
Therefore, in this model we have $N=1$ supersymmetry.

\section{Conclusion}

In this paper we have constructed asymmetric orbifold models which are based
on non-supersymmetric heterotic strings.
We have shown that a simple condition in terms of fermion currents is
necessary and sufficient for our class of asymmetric orbifolds to be
interpreted as the toroidal compactifications of non-supersymmetric heterotic
strings.
We have made a systematic classification of the asymmetric \ZN -orbifold
models with standard embeddings and obtained many asymmetric orbifold models
which are geometrically interpreted as toroidal compactifications of
non-supersymmetric strings.
We have also discussed the supersymmetric asymmetric orbifold
models constructed from non-supersymmetric heterotic strings.
We have investigated the condition for obtaining supersymmetric \ZN -orbifold
models from non-supersymmetric heterotic strings
and have shown that in order to obtain supersymmetric models we must consider
asymmetric orbifold models with non-standard embeddings.
We have constructed the examples of the $N=1$, $N=2$ and $N=4$ asymmetric
orbifold models of non-supersymmetric heterotic strings.
It would be of interest to investigate other examples of such supersymmetric
models constructed from non-supersymmetric heterotic strings.

\bigskip

\begin{center}
{\large\bf Acknowledgements}
\end{center}

We would like to thank M. Sakamoto for reading the manuscript and useful
comments.

\newpage
\pagestyle{empty}
\begin{table}
\caption{List of asymmetric \ZN -orbifold models of non-supersymmetric
heterotic strings with the right-moving twist-untwist intertwining currents.}
\vspace{3 mm}
\centering
\begin{tabular}{|c|c|c|}
\hline
Right-moving \KM\ algebra & Fermion current & \begin{tabular}{l} Number
of \\ models \end{tabular} \\
\hline
$ (SU(2))^8 $&$ SO(4) $&    4 \\
$ (SU(2))^6 \times SU(3) $&$    SO(4) $&    11 \\
$ (SU(2))^6 \times (U(1))^3 $&$ SO(2) $&    3 \\
$ (SU(2))^5 \times SU(4) $&$    SO(4) $&    4 \\
$ (SU(2))^5 \times (U(1))^2 $&$ SO(6) $&    1 \\
$ (SU(2))^4 \times SU(3) $&$    SO(8) $&    1 \\
$ (SU(2))^4 \times SO(8) $&$    SO(4) $&    14 \\
$ (SU(2))^4 \times (U(1))^5 $&$ SO(2) $&    3 \\
$ (SU(2))^4 \times (U(1))^4 $&$ SO(4) $&    3 \\
$ (SU(2))^3 \times SU(3) \times (U(1))^2 $&$    SO(6) $&    3 \\
$ (SU(2))^3 \times SU(4) $&$    SO(8) $&    1 \\
$ (SU(2))^3 \times SU(6) $&$    SO(4) $&    11 \\
$ (SU(2))^2 \times (SU(3))^2 \times (U(1))^2 $&$    SO(4) $&    1 \\
$ (SU(2))^2 \times (SU(3))^2 $&$    SO(8) $&    2 \\
$ (SU(2))^2 \times SU(3) \times (U(1))^4 $&$    SO(4) $&    1 \\
$ (SU(2))^2 \times SU(4) \times (U(1))^2 $&$    SO(6) $&    1 \\
$ (SU(2))^2 \times SU(5) $&$    SO(8) $&    4 \\
$ (SU(2))^2 \times SO(8) \times (U(1))^3 $&$    SO(2) $&    2 \\
$ (SU(2))^2 \times SO(8) $&$    SO(8) $&    8 \\
$ SU(2) \times SU(3) \times SU(4) $&$   SO(8) $&    3 \\
$ SU(2) \times SU(6) $&$    SO(8) $&    7 \\
$ SU(2) \times SO(8) \times (U(1))^3 $&$    SO(4) $&    5 \\
$ SU(2) \times SO(8) \times (U(1))^2 $&$    SO(6) $&    2 \\
$ SU(2) \times SO(10) $&$   SO(8) $&    17 \\
$ (SU(3))^4 \times U(1) $&$ SO(2) $&    15 \\
$ (SU(3))^3 $&$ SO(8) $&    2 \\
$ SU(3) \times SU(5) $&$    SO(8) $&    11 \\
$ SU(3) \times SO(8) $&$    SO(8) $&    21 \\
$ SU(3) \times (U(1))^6 $&$ SO(4) $&    1 \\
$ (SU(4))^2 $&$ SO(8) $&    2 \\
$ SU(5) \times (U(1))^3 $&$ SO(6) $&    3 \\
$ SU(6) \times (U(1))^2 $&$ SO(6) $&    4 \\
$ SU(7) $&$ SO(8) $&    14 \\
$ SO(8) \times (U(1))^3 $&$ SO(6) $&    6 \\
$ SO(12) $&$    SO(8) $&    45 \\
$ E_6 $&$   SO(8) $&    47 \\
\hline
\end{tabular}
\end{table}

\end{document}